\documentstyle[12pt]{article}

\title{The Pfaff lattice and skew-orthogonal
polynomials\footnote{Appears in Intern. Math. Research Notices
1999}}

\author{M. Adler\thanks{Department of Mathematics,
Brandeis University, Waltham, Mass 02454, USA. E-mail:
adler@math.brandeis.edu. The support of a National Science
Foundation grant \# DMS-98-4-50790 is gratefully
acknowledged.}~~~~~~E. Horozov\thanks{Department of Mathematics,
Sofia University, Bulgaria. E-mail: horozov@fmi.uni-sofia.bg. The hospitality
of the University of Louvain and Brandeis University is gratefully
acknowledged. Partially supported by Grant MM-523 of The Bulgarian Ministry of
Education.}~~~~~~P. van Moerbeke\thanks{Department
of Mathematics, Universit\'e de Louvain, 1348 Louvain-la-Neuve,
Belgium and Brandeis University, Waltham, Mass 02454, USA. E-mail:
vanmoerbeke@geom.ucl.ac.be and @math.brandeis.edu. The  support of
a National Science Foundation grant \# DMS-98-4-50790, a Nato, a
FNRS and a Francqui Foundation grant is gratefully acknowledged.}}

\date{October 28, 1998}

\newcommand{\MAT}[1]{\left(\begin{array}{*#1c}}
\newcommand{\mat}{\end{array}\right)}
\newcommand{\qed}
{%
\mbox{}%
\nolinebreak%
\hfill%
\rule{2mm}{2mm}%
\medbreak%
\par%
}

\newcommand{\sumbis}[2]%
{%

\begin{array}[t]{c}
\sum\\
{\scriptstyle #1}\\
{\scriptstyle #2}
\end{array}

}

\newcommand{\lrg}{\longrightarrow}

\newcommand{\TT}{\tilde\tau}
\newcommand{\DR}{{\cal D}}

\newcommand{\BC}{{\Bbb C}}

\newcommand{\Bk}{{\Bbb k}}
\newcommand{\Bn}{{\Bbb n}}

\newcommand{\iy}{\infty}
\newcommand{\pl}{\partial}
\newcommand{\al}{\alpha}
\newcommand{\proof}{\underline{\sl Proof}: }
\newcommand{\remark}{\underline{\sl Remark}: }

\newcommand{\HR}{{\cal H}}
\newcommand{\JR}{{\cal J}}

\newcommand{\NR}{{\cal N}}

\newcommand{\GR}{{\cal G}}
\newcommand{\vp}{\varphi}
\newcommand{\la}{\langle}
\newcommand{\ra}{\rangle}

\newcommand{\dt}{\delta}

\newcommand{\vr}{\varepsilon}
\newcommand{\sg}{\sigma}

\newcommand{\Lb}{\Lambda}
\newcommand{\tr}{\mbox{tr}}

\def\be{\begin{equation}}
\def\ee{\end{equation}}
\def\bea{\begin{eqnarray}}
\def\eea{\end{eqnarray}}

\ifx\undefined\Bbb
        \let\Bbb\bf
\fi

\catcode`\@=11
\def\ps@X{\let\@mkboth\@gobbletwo
        \def\@oddhead{\tt A - H
        -~v M:%
        ~~Skew-Orthogonal\hfil  oct 28, 1998
         \#1\hfil\S\thesection,
p.\thepage
        }
        \def\@oddfoot{\rm\hfil\thepage\hfil}
        \let\@evenhead\@oddhead
        \let\@evenfoot\@oddfoot}
\catcode`@=12
\catcode`\@=11
\let\c@equation=\relax
\newcounter{equation}[subsection]

\catcode`\@=12

\newtheorem{definition}{Definition}[
section]

\newtheorem{theorem}[definition]{Theorem}

\newtheorem{lemma}[definition]{Lemma}
\newtheorem{corollary}[definition]{Corollary}
\newtheorem{proposition}[definition]{Proposition}

\catcode`\@=11
\let\c@equation=\relax
\newcounter{equation}[
section]

\catcode`\@=12

\begin{document}
\maketitle

\abstract Consider a semi-infinite skew-symmetric moment matrix,
$m_{\iy}$ evolving according to the vector fields $\pl m / \pl
t_k=\Lb^k m+m \Lb^{\top k} ,$ where $\Lb$ is the shift matrix. Then
The skew-Borel decomposition $ m_{\iy}:= Q^{-1} J Q^{\top -1} $
leads to the so-called Pfaff Lattice, which is integrable, by
virtue of the AKS theorem, for a splitting involving the affine
symplectic algebra. The tau-functions for the system are shown to
be pfaffians and the wave vectors skew-orthogonal polynomials; we
give their explicit form in terms of moments. This system plays an
important role in symmetric and symplectic matrix models and in the
theory of random matrices (beta=1 or 4).

\setcounter{equation}{0}

\vspace{0.5cm}

\tableofcontents


\vspace{1cm}

\noindent {\bf Lie algebra splitting}. Throughout this paper the Lie algebra $\DR = gl_{\iy}$ of
semi-infinite matrices is viewed as composed of $2\times 2$ blocks.
It admits the natural decomposition into subalgebras:
\be
\DR=\DR_-\oplus\DR_0\oplus\DR_+=\DR_-\oplus\DR_0^-\oplus\DR_0^+\oplus\DR_+
\ee
where $\DR_0$ has $2\times 2$ blocks along the diagonal with zeroes
everywhere else and where $\DR_+$ (resp. $\DR_-$) is the subalgebra
of upper-triangular (resp. lower-triangular) matrices with $2
\times 2$ zero matrices along $\DR_0$ and
zero below (resp. above). As pointed out in (0.1), $\DR_0$ can
further be decomposed into two Lie subalgebras:
\bea
\DR_0^-&=&\{\mbox{all $2\times 2$ blocks $\in \DR_0$
are proportional to Id}   \}\nonumber\\
\DR_0^+&=&\{\mbox{all $2\times 2$ blocks $\in \DR_0$
have trace $0$ }  \}.
\eea
Consider the semi-infinite skew-symmetric matrix $J$, zero
everywhere, except for the following $2\times 2$ blocks, along the
``diagonal",
\be
J=\pmatrix{0&1& & & & &\cr
-1&0& & & & & \cr
 & & 0&1& & &\cr
 & & -1&0& & &\cr
 & & & & 0&1 &\cr
 & & & & -1&0 &\cr
 &&&&&&\ddots }\in \DR^+_0,~~\mbox{with}~J^2=-I,
\ee
and the associated Lie algebra order 2 involution
\be\JR:\DR\longrightarrow
\DR: a\longmapsto\JR(a):=Ja^{\top}J.
\ee
The splitting into two Lie subalgebras\footnote{Note $\Bn$ is the
fixed point set of $\JR$.}
\be
\DR= \Bk+\Bn,~~\mbox{with} ~~\Bk=\DR_-+\DR_0^-~~
\mbox{and}~~ \Bn= \{a+\JR a,~a
\in \DR  \}=\mbox{sp}(\iy),
\ee
with corresponding Lie groups\footnote{$\GR_{\Bk}$ is the group of
invertible elements in $\Bk$, i.e., lower-triangular matrices, with
non-zero $2\times 2$ blocks proportional to Id along the diagonal.}
$\GR_{\Bk}$ and $\GR_{\Bn}=Sp(\iy)$, will play a crucial role here.
Let $\pi_{\Bk}$ and $\pi_{\Bn}$ be the projections onto $\Bk$ and
$\Bn$. Notice that $\Bn=$sp$(\iy)$ and $\GR_{\Bn}=Sp(\iy)$ stand
for the infinite rank affine symplectic algebra and group; e.g. see
\cite{Kac}; the exposition in this paper is totally self-contained
and does not depend on any knowledge of the affine symplectic
algebra.

\medbreak

\noindent{\bf Pfaff Lattice}. As will be shown in this
paper, the Lax pair, which we call the Pfaff lattice\footnote{The
Hamiltonians $\HR_i$ are viewed as formal sums; the convergence of
this formal sum would require some sufficiently fast decay of the
entries of $L$. Since $\nabla
\HR_i=L^i$, one does not need to be concerned about this point.},
\be
\frac{\pl L}{\pl t_i}=\left[-\pi_{\Bk}\nabla \HR_i, L \right]=
\left[\pi_{\Bn}\nabla \HR_i, L \right],
~~\mbox{with}~~  \HR_i=\frac{\tr L^{i+1}}{i+1},
\ee
\hfill ({\bf Pfaff Lattice})

\noindent on matrices $L=Q\Lb Q^{-1}$, with $Q\in \GR_{\Bk}$ and $\Lb$ the
customary shift operator, is completely integrable, as a result of
the AKS-theorem. The Lax pair, written in compact form in (0.6), is
given explicitly in (1.4). In \cite{ASV}, this lattice was
investigated from the point of view of the 2d-Toda
 lattice with special initial conditions. In
\cite{AvM5}, we developped the wave- and tau-function theory for
this lattice and we exhibited a map from the Toda to the Pfaff
lattice.

\medbreak

\noindent{\bf Linearizing vector fields and Pfaffian $\TT$-functions}. The key to this system is its linearization (moment map):
  namely, if $L=Q\Lb Q^{-1}$ flows according to (0.6), then the
following skew-symmetric matrix, constructed from $Q$,
$$
m_{\iy}:= Q^{-1} J Q^{\top -1}
$$
flows as
 \be
 \frac{\pl m_{\iy} }{ \pl t_k}=\Lb^k m_{\iy}
+m_{\iy} \Lb^{\top k}  ~~, k=1,2,3,...~.
\ee
Having constructed the skew-symmetric matrix $m_{\iy}(t)
 =\left(
\mu_{ij}(t)\right)_{0 \leq i,j \leq \iy}$,
 we define the skew-symmetric submatrices
\be
m_{n}(t)=\left( \mu_{ij}(t)\right)_{0 \leq i,j \leq n-1}  ,
~~\mbox{with}~~ \TT_{2n}(t)={\det}^{1/2} m_{2n}(t),~~t \in
\BC^{\iy}.
\ee
Note that, since $m_n$ is skew-symmetric, $\det m_{2n+1}=0$,
whereas $\det m_{2n}$ is a perfect square, with ${\det}^{1/2}
m_{2n}$ being the Pfaffian
; for a definition, specifying the sign, see section 3.
 The technology of letting $m_{\iy}$ flow in time
 and Borel decomposing $m_{\iy}$
has been used quite extensively in other situations, like the
standard Toda lattice or the 2-Toda lattice \cite{ASV}.

 The so-called {\em Pfaffian} $\TT$-{\em function} is
not a KP $\tau$-function, but enjoys different bilinear identities
and Hirota-type bilinear equations \cite{ASV},
\bigbreak

$
\{\TT_{2n}(t-[u]),\TT_{2n}(t-[v])\}
$\hfill
\bea
& & \hspace{0cm}+(u^{-1}-v^{-1})(\TT_{2n}(t-[u])\TT_{2n}(t-[v])
-\TT_{2n}(t)\TT_{2n}(t-[u]-[v]))\nonumber\\
 & &=uv(u-v)\TT_{2n-2}(t-[u]-[v])\TT_{2n+2}(t)
 ,~\mbox{where}
 ~[u]=(u,\frac{u^2}{2},...).\nonumber
\eea
In particular,
 \be
\left(p_{k+4}(\tilde\pl)-\frac{1}{2}\frac{\pl^2}{\pl
t_1\pl t_{k+3}}\right)\TT_{2n}\circ\TT_{2n}=p_k(\tilde
\pl)~\TT_{2n+2}\circ\TT_{2n-2},
\ee
for $k,n=0,1,2,...~.$ Both these equations are reminiscent of the
differential Fay identities and the KP equations for usual
$\tau$-functions \cite{Dickey,vM}, but with an additional term on
the right hand side, involving $\TT_{2n+2}$ and $\TT_{2n-2}$.

For $k=0$, this KP-like equation has already appeared in the
context of the charged BKP hierarchy, studied by V. Kac and van de
Leur \cite{KvdL}; the precise relationship between the charged BKP
hierarchy of Kac and van de Leur and the Pfaff Lattice, introduced
here, deserves further investigation.

\medbreak

\noindent {\bf Skew-Borel decomposition and skew-orthogonal polynomials}.
Starting with equations (0.7), we show they have a unique
semi-infinite skew-symmetric matrix solution $m_{\iy}(t)$, provided
the initial
 condition satisfies $\det m_{2n}\neq 0$, for $n=0,1,2,..$. It is given in terms
 of its (unique)
``skew Borel decomposition"\footnote{unique, modulo a sequence of
$\pm$-signs.}
\be
m_{\iy}=Q^{-1} J Q^{\top -1},~~\mbox{with}~~Q \in \GR_{\Bk}.
\ee
Here, we view $m_{\iy}$ as a matrix, providing the ``{\em
inner-product}" between monomials,
 \be
 \la y^i, z^j \ra:= (m_{\iy})_{ij}=:\mu_{ij}
 .\ee
Then the lower-triangular matrix Q is conveniently described in
 terms of a sequence of so-called "{\em skew-orthonormal
 polynomials}"\footnote{$\chi(z)=(1,z,z^2,...)$.}
$$q(z)=(q_0(z),q_1(z),...)^{\top}=Q \chi(z)~\mbox{satisfying}
~
\left( \la q_i, q_j\ra\right)_{i,j \geq 0} =J.
$$
Skew-orthonormal
  polynomials were first introduced by Mehta \cite{M}; see also Br\'ezin and Neuberger
  \cite{BN}, where they appeared in the context of unoriented
  random surfaces. To the best of our knowledge, neither the form, nor the connection with the
  Pfaff lattice was known.

We show the polynomials $q(t,z)$ in $z$, depending on $t$, are
explicitly given by pfaffians of skew-symmetric matrices, which in
the even case ($q_{2n}$'s) are formed
 by
replacing in $m_{2n+2}$ the $2n+2$th row and column by powers of
$z$ and in the odd case ($q_{2n+1}$'s), by replacing the $2n+1$th
row and column by the same powers of $z$, keeping the skewness of
the matrices:
\bigbreak

\noindent $q_{2n}(t,z)$
\bea
&= &\left( \TT_{2n} \TT_{2n+2} \right)^{-1/2}
 pf
\pmatrix{0&
\mu_{01}& ~~...~~&\mu_{0,2n}&1\cr
-\mu_{01}& 0 & ~~...~~& \mu_{1,2n}&z\cr
\vdots& \vdots &   & \vdots&\vdots\cr
-\mu_{0,2n}&-\mu_{1,2n} & ~~...~~&0 &z^{2n} \cr
-1& -z & ~~...~~&-z^{2n}&0\cr}\nonumber\\
&&\nonumber\\\nonumber
\eea

\bigbreak

\noindent $q_{2n+1}(t,z)$
\bea
 &&\nonumber\\ &&\nonumber\\
 &=&\left( \TT_{2n} \TT_{2n+2}
\right)^{-1/2} pf
\pmatrix{0&
\mu_{01}& ~~...~~&1&\mu_{0,2n+1}\cr
-\mu_{01}& 0 & ~~...~~&z& \mu_{1,2n+1}\cr
\vdots& \vdots &   & \vdots&\vdots\cr
-1& -z & ~~...~~&0&-z^{2n+1}\cr
-\mu_{0,2n+1}&-\mu_{1,2n+1} & ~~...~~&z^{2n+1} &0 \cr}\nonumber\\
 &=&(\TT_{2n}\TT_{2n+2})^{-1/2}
 \left( z+\frac{\pl}{\pl t_1}\right)(\TT_{2n}\TT_{2n+2})^{1/2}
 q_{2n}(t,z)
. \nonumber\\
\eea
Finally, the matrix
\be
L=Q\Lb Q^{-1},~~\mbox{satisfying}~~ z q(t,z)= L(t) q(t,z),
\ee
forms a unique semi-infinite solution to the Lax pair (0.6).

It is interesting to point out the striking similarity with the
orthogonal polynomials, the Toda lattice and the matrix models,
described in \cite{AvM1}. Both, standard Toda\footnote{Isospectral
deformations on tridiagonal symmetric matrices.} and Pfaff lattices
can be viewed as reductions of the 2-Toda lattice, where the
initial condition $m_{\iy}$ is a H\"ankel matrix for the standard
Toda lattice and a skew-symmetric matrix for the Pfaff lattice. In
the former case, the motion takes place within the big stratum and
in the latter, within a deeper stratum.

\medbreak

\section{ The Pfaff Lattice and a Lie algebra splitting}

Remember the decomposition (0.1) of the Lie algebra into
subalgebras:
\be
\DR=gl_{\iy}=\DR_-\oplus\DR_0\oplus\DR_+=\DR_-\oplus\DR_0^-\oplus\DR_0^+\oplus\DR_+
.\ee
Remember the element $J \in \DR_0^+$, with $J^2=-I$ and the
associated map:
\be
\JR:\DR\longrightarrow \DR:  a\longmapsto\JR(a):=Ja^{\top}J.
\ee

\begin{theorem}
The vector fields\footnote{$\nabla$ denotes the gradient in the Lie
algebra $\DR $ with respect to the natural pairing $\DR$ with
$\DR^*=\DR $, i.e., $\nabla
\HR_i=L^i$.}
\be
\frac{\pl L}{\pl t_i}=\left[-\pi_{\Bk}\nabla \HR_i, L \right]=
\left[\pi_{\Bn}\nabla \HR_i, L \right]
~~\mbox{with}~~\HR_i=\frac{\tr L^{i+1}}{i+1},
\ee
or written out as\footnote{$\JR (L^i)_+$ means $\JR
 \left((L^i)_+\right);$ the same for $\JR (L^i)_0$.}
\bea
\frac{\pl L}{\pl
t_i}&=&\left[-\left((L^i)_--\JR (L^i)_+\right)-\frac{1}{2}
\left((L^i)_0-\JR (L^i)_0 \right),L\right]\nonumber\\
 &=&\left[\left((L^i)_++\JR (L^i)_+\right)+\frac{1}{2}
\left((L^i)_0+ \JR (L^i)_0 \right),L\right],
\eea
all commute.
\end{theorem}

The proof of this theorem hinges on a number of Lemmas and the
Adler-Kostant-Symes theorem \cite{AvM0}:

\begin{lemma} $\JR$ is a Lie algebra isomorphism and an order 2 involution on $\DR$:
\be
\JR^2=I,
\ee
with $\NR_+$ and $\NR_-$ being the $\pm 1$ eigenspaces of $\JR$:
\be
\NR_{\pm}=\{\mbox{$a$ such that $\JR a=\pm a\}=\{b\pm\JR b$ with
$b\in\DR\}$}.
\ee
Then $\DR$ is decomposed into the Lie algebra $\NR_+$ and the
(symmetric) vector space $\NR_-$
\be
\DR=\NR_++\NR_-,\mbox{ with }[\NR_{\pm},\NR_{\pm}]
\subset \NR_+\mbox{ and }[\NR_+,\NR_-]\subset \NR_-,
\ee
with
\be
\NR_{\pm}=\DR^{\pm}_0\oplus(\NR_{\pm}\backslash\DR_0),
\ee
and
$$
\NR_{\pm}\backslash\DR_0=\{a\pm \JR a \mbox{ such that }
 a\in\DR_-~\mbox{or}~\DR_+\}
$$
\be
\DR_0^{\pm}=\DR_0\cap \NR_{\pm}=\{a\pm \JR a \mbox{ such that }a\in\DR_0\}.
\ee
\end{lemma}

\proof The map $\JR$ is a Lie algebra homomorphism,
 such that $\JR^2=I$; indeed, upon using $J^2=-I$,
$$\JR([a,b])=J[a,b]^{\top}J=-J[a^{\top},b^{\top}]J
=[Ja^{\top}J,Jb^{\top}J]=[\JR a,\JR b].
$$
In particular, $\JR(a)=\pm a$ and $\JR(b)=\pm b$ imply
$$
\JR[a,b]=[\pm a,\pm b]=[a,b],
$$
leading to the inclusions (1.7); so $\NR_+$ is a Lie subalgebra of
$\DR$, namely an infinite-dimensional version of the symplectic
algebra $\mbox{sp}(\iy)$. The second description (1.6) of
$\NR_{\pm}$ follows from
$$
\JR(b\pm\JR b)=\JR b\pm\JR^2 b=\pm(b\pm\JR b),
\quad\mbox{for }b\in\DR;
$$
and the vector space decomposition $\DR=\NR_++\NR_-$:
$$a=\frac{1}{2}(a+\JR a)+\frac{1}{2}(a-\JR a).
$$
Since $J\DR_{\vr},\DR_{\vr}J \subset \DR_{\vr},
\DR_{\vr}^{\top}=\DR_{-\vr},$ for $\vr=+,-,0 $, we have
$$
\JR:\DR_{\vr}\lrg\DR_{-\vr},
$$
and so, if $a\in\DR_+$, we have $b:=\JR a\in\DR_-$ and
$$
\NR_{\pm}\supset a\pm\JR a=\JR^2 a\pm\JR a=\pm(b\pm\JR b)
\quad\mbox{with}\quad b=\JR a\in\DR_-.
$$
Therefore
$$
\{b\pm\JR b\mbox{ such that }b\in\DR_-\}=\{a\pm\JR a\mbox{ such that }
b\in\DR_+\}=\NR_{\pm}\backslash\DR_0
$$
establishing the first relation (1.9). As to the second, one checks
that
\be
\MAT{2}a&b\\c&d\mat\pm\MAT{2}0&1\\-1&0
\mat\MAT{2}a&b\\c&d\mat^{\top}
\MAT{2}0&1\\-1&0\mat =\left\{
\begin{array}{c}
\MAT{2}a-d&2b\\
2c&-(a-d)\mat\\
 \\
(a+d)I
\end{array}
\right.\,.
\ee
\qed

\begin{lemma} The decomposition
$$\DR=\Bk\oplus\Bn,
$$
with
$$
\Bk=\DR_-\oplus\DR_0^-,\quad \Bn=\NR_+=
\{ a ~\mbox{such that}~\JR a=a \}=\mbox{sp}(\iy),$$
is a vector space splitting of $\DR$ into Lie subalgebras. Any
element $a\in\DR$ decomposes uniquely into its projections onto
$\Bk$ and $\Bn$:
\bea
a&=&\pi_{\Bk}a+\pi_{\Bn}a\nonumber\\ &=&\left\{\left(a_--\JR
a_+\right)+\frac{1}{2}\left(a_0-\JR a_0
\right)\right\}+\left\{\left(a_++\JR a_+\right)+\frac{1}{2}\left(a_0+\JR
a_0
\right)\right\}\nonumber.
\eea
\end{lemma}

\proof  Since $[\DR_-,\DR_-]\subset\DR_-$, since
 $\DR_-\DR_0, ~\DR_0\DR_-\subset\DR_-$ and $[\DR^-_0,\DR^-_0]=0$, we have that
 $[\Bk,\Bk]=\DR_-\subset\Bk$, so that
$\Bk =\DR_-+\DR^-_0$ and
$\Bn=\NR_+$ are both Lie subalgebras.

But also $\Bk \cap \Bn =0$; indeed, a typical element of $\NR_+$
has the form,
$$
(b_-+\JR(b_-))+(d_0+\JR(d_0))\in(\NR_+\backslash\DR_0)+(\NR_+\cap\DR_0)=
(\NR_+\backslash\DR_0)+\DR_0^+,
$$
with $b_-\in\DR_-$, $d_0\in\DR_0$. If this element
$\in\Bk=\DR_-+\DR^-_0$, then $\JR(b_-)=0$ and thus $b_-=0$, but
then, since $d_0+\JR d_0 \in \DR_0^+ \cap
\DR_0^-=0 $. Finally, any element $a \in \DR$ can be decomposed
into\footnote{$a_{\pm}=\pi_{\DR_{\pm}}a,~a_{0}=\pi_{\DR_{0}}a.$}
\bea
a&=&a_-+a_0+a_+\nonumber\\ &=&\left\{\left(a_--\JR
a_+\right)+\frac{1}{2}\left(a_0-\JR a_0
\right)\right\}\nonumber\\
& &\quad +\left\{\left(a_++\JR a_+ \right)
+\frac{1}{2}\left(a_0+\JR
a_0
\right)\right\}\nonumber\\
&\in&(\DR_-+\DR_0^-)\oplus\left((\NR_+\backslash\DR_0)+\DR^+_0\right)=\Bk\oplus\Bn.\nonumber
\eea
\qed

We now state the R-matrix version of the AKS-theorem (see
\cite{AvM0} and \cite{RS}). The substance of the R-matrix extension
is that initial conditions $\xi(0)$ can be taken in all of ${\bf
g}$, instead of merely in ${\bf k}$.

\begin{proposition} {\bf (Adler-Kostant-Symes, Russian flavor)}  Let ${\bf g} = {\bf k} + {\bf n}$ be a
(vector space) direct sum of a Lie algebra  ${\bf g}$ in terms of
Lie subalgebras  ${\bf k}$ and  ${\bf n}$, with ${\bf g}$ paired
with itself via a non-degenerate {\rm ad}-invariant inner
product\footnote{$\la{\rm Ad}_gX;Y\ra =\la X,{\rm Ad}_{g-1}Y\ra$,
$g\in G$, and thus $\la [z,x],y\ra =\la x,-[z,y]\ra.$} $\la\,
,\,\ra$; this in turn induces a decomposition ${\bf g} = {\bf
k}^{\bot} + {\bf n}^{\bot}$ and isomorphisms  ${\bf g}\simeq {\bf
g}^*$, ${\bf k}^{\bot}\simeq {\bf n}^*$, ${\bf n}^{\bot}\simeq {\bf
k}^*$.  It also leads to a second Lie algebra  ${\bf g}_R
\simeq {\bf g}_R^*$ derived from ${\bf g}$, namely:
$$ {\bf g}_R : [x,y]_R = \frac{1}{2}[Rx,y] +
\frac{1}{2}[x,Ry] = [\pi_{{\bf k}}x,\pi_{{\bf k}}y] -
[\pi_{{\bf n}}x,\pi_{{\bf n}}y],  $$ with  $R =
\pi_{{\bf k}}-\pi_{{\bf n}}$.
 {\rm Ad}$^* \simeq$ {\rm Ad}-invariant functions $\vp $
 on ${\bf g}^* \simeq {\bf g}$ Poisson commute for
  the
Kostant-Kirillov  Poisson structure
$$
\{f,h\}_R (\xi) :=\la \xi,[\nabla h(\xi),
\nabla f(\xi)]_R\ra
~~\mbox{ on} ~~{\bf g}^*_R\simeq{\bf g}_R.
$$
 The associated
Hamiltonian flows are expressed in terms of the Lax
pair\footnote{$\nabla\vp$ is defined as the element in
${\bf g}^*$
such that $d\vp(\xi)=\la\nabla\vp,d\xi\ra$,
$\xi\in{\bf g}$.}
$$
\dot\xi =[-\pi_{{\bf
k}}\nabla\vp(\xi),\xi]=[\pi_{{\bf
n}}\nabla\vp(\xi),\xi]~~\mbox{\,for\,}~~\xi\in{\bf g}_R.
$$
The systems has a solution expressible in two different
ways\footnote{naively written Ad$_{K(t)}\xi_0=K(t)\xi_0K(t)^{-1}$,
Ad$_{N^{-1}}\xi_0=N^{-1}(t)\xi_0N(t)$.}:
\be
\xi(t)={\rm Ad}_{K(t)}\xi_0={\rm Ad}_{N^{-1}(t)}\xi_0
\ee
with\footnote{with regard to the group factorization
$A=\pi_{\GR_{\Bk}}A~
\pi_{\GR_{\Bn}}A$.}
$$
K(t) = \pi_{\GR_{\Bk}} e^{t \nabla\vp(\xi_0)},\quad
 \mbox{and}\quad
N(t)
=
 \pi_{\GR_{\Bn}} e^{t \nabla\vp(\xi_0)}.
$$
\end{proposition}

\noindent \underline{{\em Proof of Theorem 1.1}}: Identifying
$\Bk$ and $\Bn$ of Lemma 1.3, with those of proposition 1.4 and
unraveling the projection $\pi_{\Bk}$, according to Lemma 1.3,
imply that the vector fields (1.3) and so (1.4) all commute.\qed

\section{The vector fields $\pl m / \pl t_k=\Lb^k m
+m \Lb^{\top k} $}

The main claim of this section can be summarized in the following
statement:

\begin{theorem}
Consider the skew-symmetric solution
 $$m_{\iy}(t)=e^{\sum t_k \Lb^k}m_{\iy}(0)e^{\sum t_k \Lb^{\top k}}$$
 to the commuting equations
\be
\frac{\pl m_{\iy} }{ \pl t_k}=\Lb^k m_{\iy}
+m_{\iy} \Lb^{\top k}, \ee
 with skew-symmetric initial condition $m(0)$ and
 its ``skew-Borel decomposition"
\footnote{Remember $\GR_{\Bk}$ denotes the group of invertible elements
 in $\Bk$.}
\be
m_{\iy}=Q^{-1} J Q^{-1 \top}, ~~\mbox{with}~~Q \in \GR_{\Bk}.
\ee
Then the matrix $Q$ evolves according to the equations
\be
\frac{\pl Q}{\pl t_i}Q^{-1} =-\pi_{{\bf k}}
\left( Q \Lb^i Q^{-1} \right)
\ee
and the matrix $L:=Q \Lb Q^{-1}$ provides a solution to the Lax
pair
\be
\frac{\pl L}{\pl t_i}=\left[-\pi_{{\bf k}} L^i,L    \right]
=\left[\pi_{{\bf n}} L^i,L    \right].
\ee
Conversely, if $Q \in \GR_{\Bk}$ satisfies (2.3),
 then $m_{\iy}$, defined by (2.2),
satisfies (2.1).
\end{theorem}

The proof of this theorem hinges on the following proposition:

\begin{proposition} For the matrices
$$
L:=Q\Lb Q^{-1}\quad\mbox{and}\quad m:=Q^{-1}J Q^{-1\top}, \mbox{
with }Q\in \GR_{\Bk},
$$
the following three statements are equivalent\medbreak\indent (i) $\displaystyle{\frac{\pl Q}{\pl
t_i}Q^{-1}=-\pi_{\Bk}L^i}$
\medbreak\indent (ii) $L^i+\displaystyle{\frac{\pl Q}{\pl
t_i}}Q^{-1}\in\Bn$\medbreak\indent (iii) $\displaystyle{\frac{\pl m}{\pl
t_i}}=\Lambda^im+m\Lambda^{\top^i}.$
\medbreak\noindent Whenever the vector fields on $Q$ or $m$
satisfy (i), (ii) or (iii), then the matrix $L=Q\Lb Q^{-1}$ is a
solution of the Lax pair
$$
\frac{\pl L}{\pl t_i}=[-\pi_{\Bk}L^i,L]=
 \left[\pi_{{\bf n}} L^i,L    \right].
$$
\end{proposition}

\proof Written out, proposition 2.2 amounts to showing the
equivalence of the three formulas:
\medbreak
\noindent(I) $\displaystyle{\frac{\pl Q}{\pl t_i}Q^{-1}+\left((L^i)_--J(L^i_+)^{\top}J\right)+\frac{1}{2}
\left((L^i)_0-J((L^i)_0)^{\top}J\right)=0}$

\medbreak

\noindent(II) $\displaystyle{\left(L^i+\frac{\pl Q}{\pl t_i}Q^{-1}\right)-J
\left(L^i+\frac{\pl Q}{\pl t_i}Q^{-1}\right)^{\top}J=0}$

\medbreak

\noindent(III) $\Lambda^im+m\Lambda^{\top i}-\displaystyle{\frac{\pl m}{\pl
t_i}}=0$.

\medbreak

\noindent The point is to show that
$$
(\mbox{I})_+=0,~~~~(\mbox{I})_-=(\mbox{II})_-
=-J~(\mbox{II}_+)^{\top} ~J,~~~~
(\mbox{I})_0=\frac{1}{2}\left(\mbox{II}\right)_0,
$$
\be
Q^{-1}(\mbox{II})JQ^{-1^{\top}}
=(\mbox{III}).
\ee
\noindent The fact that $Q\in\GR_{\Bk}\subset
\Bk=\DR_-\oplus\DR_0^-=\DR_-\oplus(\DR_0 \cap \NR^-)
 $ amounts to
$$
Q_+=0\quad\mbox{and}\quad(Q+JQ^{\top}J)_0=0.
$$
The latter statement implies that $Q$ is lower-triangular and,
along the diagonal, is composed of $2\times 2$ blocks proportional
to the identity. Hence
$$
Q_0,\dot Q_0\in\DR_0^-,\quad\mbox{and so }~~(\dot Q
Q^{-1})_0\in\DR_0^-
$$
and thus $J$ commutes with $(\dot Q Q^{-1})_0$, yielding:
\be
J(\dot Q Q^{-1})_0J=J^2(\dot Q Q^{-1})_0=-(\dot Q Q^{-1})_0.
\ee
Also notice that for any matrix $A$
$$
(JAJ)_{\pm}=JA_{\pm}J\quad\mbox{and}\quad  (JAJ)_0=JA_0J.
$$
At first, one observes that
\be
(\mbox{I})_+=0.
\ee
Also
\bea
(\mbox{II})_-&=&\left(L^i+\dot Q Q^{-1}-J(L^i+\dot Q
Q^{-1})^{\top}J\right)_-\nonumber\\ &=&\left(L^i+\dot Q
Q^{-1}-J((L^i)_+)^{\top}J\right)_-,\quad\mbox{using }(\dot Q
Q^{-1})_+=0\nonumber\\ &=&(\mbox{I})_-
\eea

\bea
(\mbox{II})_0&=&\left(L^i+\dot Q Q^{-1}-J\left(L^i+\dot Q
Q^{-1}\right)^{\top}J\right)_0\nonumber\\
&=&\left(L^i\right)_0-J\left(L^{i\top}\right)_0 J+\left(\dot Q
Q^{-1}\right)_0-J\left((\dot Q Q^{-1})^{\top}\right)_0 J\nonumber\\
&=&(L^i)_0-J(L^{i
\top})_0 J+2(\dot Q Q^{-1})_0,\quad\mbox{using (2.6)}\nonumber\\
&=&2(\mbox{I})_0
\eea
and
\bea
(\mbox{II})_+&=&\left(L^i+\dot Q Q^{-1}-J(L^i+\dot Q
Q^{-1})^{\top}J\right)_+\nonumber\\
&=&(L^i)_+-J((L^i)_-)^{\top}J-J((\dot Q
Q^{-1})_-)^{\top}J\nonumber\\
&=&\left(((L^i)_+)^{\top}-J(L^i)_-J-J(\dot Q
Q^{-1})_-J\right)^{\top}\nonumber\\
&=&J\left(J((L^i)_+)^{\top}J-J^2(L^i)_-J^2-J^2(\dot Q
Q^{-1})_-J^2\right)^{\top}J\nonumber\\ &=&-J\left((L^i+\dot Q
Q^{-1}-J(L^i_+)^{\top}J)_-\right)^{\top}J\nonumber\\
&=&-J((\mbox{I})_-)^{\top}J.
\eea

\noindent Using the definitions $L=Q\Lambda Q^{-1}$, $m=Q^{-1}JQ^{-1\top}$
and $J^2=-I$, one finds
\bea
(\mbox{III})&=&\Lambda^im+m\Lambda^{\top^{i}}-\frac{\pl m}{\pl
t_i}\nonumber\\ &=&\Lambda^im+m\Lambda^{\top^{i}}+(Q^{-1}\dot
Q)m+m(Q^{-1}\dot Q)^{\top},\mbox{ setting
}m=Q^{-1}JQ^{-1^{\top}}\nonumber\\ &=&Q^{-1}(Q\Lambda^i
Q^{-1})JQ^{-1^{\top}}-Q^{-1}(JQ^{-1^{\top}}
\Lambda^{{\top}^i}Q^{\top}J)JQ^{-1^{\top}}\nonumber\\ & &\quad +\,Q^{-1}(\dot
QQ^{-1})JQ^{-1^{\top}}-Q^{-1}\left(JQ^{-1^{\top}}\dot
Q^{\top}J\right)JQ^{-1^{\top}}\nonumber\\ &=&Q^{-1}\left(L^i+\dot
QQ^{-1}-J(L^i+\dot QQ^{-1})^{\top}J\right)JQ^{-1^{\top}}\nonumber\\
&=&Q^{-1}(\mbox{II})JQ^{-1^{\top}}.
\eea
The five facts (2.7) up to (2.11) imply (2.5) and so the desired
equivalences.\qed

\section{Skew-orthogonal polynomials}

As before, consider the skew-symmetric solution
$$m_{\iy}(t)=e^{\sum t_k \Lb^k}m_{\iy}(0)
e^{\sum t_k \Lb^{\top k}}$$ to the commuting equations (2.1). We
now perform the ``{\em Borel decomposition}" of $m_{\iy}$:
\be
m_{\iy}=Q^{-1} J Q^{\top -1},
\ee
where $Q\in \GR_{\Bk}$ (lower-triangular invertible matrices in
$\Bk$). To carry out this factorization, we found the following
useful algorithm, based on the observation that the skew-Borel
decomposition is tantamount to finding skew-orthogonal polynomials,
in the same way that Borel decomposing a H\"ankel matrix is
tantamount to finding orthogonal polynomials.

Define the skew-inner product $\la f,g\ra$ between formal Taylor
series,
 induced by
$$\la y^i,z^j\ra :=\mu_{ij}(t)=(m_{\iy}(t))_{ij}.$$
 Monic polynomials
$p_k$ are said to be {\em skew-orthogonal}, when \be(\la
 p_i,
p_j\ra)_{0\leq i,j<\iy}=J\tilde h, ~~\mbox{with}~~
\tilde h\in
\DR_0^-.\ee
They, in turn, lead to {\em
 skew-orthonormal} polynomials $q_k(z)$,
 \be
q_{2n}(t,z)=\frac{1}{\sqrt{\tilde h_{2n}}}p_{2n}(t,z)~~\mbox{and}~~
q_{2n+1}(t,z)=\frac{1}{\sqrt{\tilde h_{2n}}}p_{2n+1}(t,z),\ee
 thus satisfying
\be(\la q_i,q_j\ra)_{0\leq i,j<\iy}=J.\ee

Indeed, finding a set of skew-orthonormal polynomials is tantamount
to the ``{\em skew-Borel decomposition}" (3.1).
 Using the lower-triangular matrix
$Q$ of coefficients of the polynomials $q(z)=Q
\chi(z)$, we have

\be J=\left(\la q_i,q_j   \ra  \right)_{i,j\geq 0}
=\left(\la (Q\chi)_i,(Q\chi)_j   \ra  \right)_{i,j\geq 0}
=Q\left(\la y^i,z^j   \ra  \right)_{i,j\geq 0}Q^{\top}=Qm_{\iy}
 Q^{\top}
,\ee from which (3.1) follows, and vice-versa.


Given a skew-symmetric matrix
$$
A=(a_{ij})_{0\leq i,j\leq n-1},\quad\mbox{even $n$},
$$
we define the Pfaffian by means of the formula\footnote{In the formula below $
 (i_0,i_1, ...
,i_{n-2},i_{n-1})=\sigma (0,1,...,n-1)$, where $\sigma$ is a
permutation and $\vr(\sigma) $ its parity.}

\medbreak

\noindent $\displaystyle pf(A)dx_0\wedge ...\wedge dx_{n-1}$
 \hfill
 \begin{eqnarray}
&=&\frac{1}{(n/2)!}\left(
\sum_{0\leq i<j\leq n-1}
a_{ij}dx_i\wedge dx_j\right)^{n/2}\nonumber\\
&=&\frac{1}{2^{n/2}(n/2)!}\left(\sum_{\sg}\vr(\sg)a_{i_0,i_1}a_{i_2,i_3}...
a_{i_{n-2},i_{n-1}}
\right)dx_0\wedge  ...\wedge dx_{n-1},\nonumber\\
\end{eqnarray}
so that $pf(A)^2=\det A$, but with a sign specification for
$pf(A)$. Henceforth, we shall be using the notation
$$
pf(i_0,i_1,...,i_{\ell-1}),~0\leq i_0< i_1<...<i_{\ell-1},~~\ell
~\mbox{even},
$$
to denote the pfaffian of the skew-symmetric matrix, formed by the
$i_0, i_1,...,i_{\ell-1}$th rows and columns of $m_{\iy}$.

\begin{theorem}
Consider a semi-infinite skew-symmetric matrix $m_{\iy}$,
satisfying (2.1); then the monic polynomials $p_k$
\bea
p_{2n}(t,z)&=&\frac{1}{\TT_{2n}(t)}
 ~pf
\pmatrix{0&
\mu_{01}& ~~...~~&\mu_{0,2n}&1\cr
-\mu_{01}& 0 & ~~...~~& \mu_{1,2n}&z\cr
\vdots& \vdots &   & \vdots&\vdots\cr
-\mu_{0,2n}&-\mu_{1,2n} & ~~...~~&0 &z^{2n} \cr
-1& -z & ~~...~~&-z^{2n}&0\cr}\nonumber\\
&&\nonumber\\
 &&\nonumber\\ &&\nonumber\\
 p_{2n+1}(t,z)&=&\frac{1}{\TT_{2n}} ~pf
\pmatrix{0&
\mu_{01}& ~~...~~&1&\mu_{0,2n+1}\cr
-\mu_{01}& 0 & ~~...~~&z& \mu_{1,2n+1}\cr
\vdots& \vdots &   & \vdots&\vdots\cr
-1& -z & ~~...~~&0&-z^{2n+1}\cr
-\mu_{0,2n+1}&-\mu_{1,2n+1} & ~~...~~&z^{2n+1} &0 \cr}\nonumber\\
 &=&\frac{1}{\TT_{2n}(t)}
 \left( z+\frac{\pl}{\pl t_1}\right)\TT_{2n}(t)
 p_{2n}(t,z)
. \nonumber\\
\eea
form a skew-orthogonal sequence, $\la  p_i,  p_j\ra_{0\leq
i,j<\iy}=J\tilde h$, where
$$  \tilde h=(\tilde h_0,\tilde h_0,\tilde h_2,\tilde
h_2,...)\in
\DR_0^-,~~\mbox{with}~~\tilde h_{2n}=
\frac{\tilde \tau_{2n+2}}{\tilde \tau_{2n}},
~~\tilde \tau_{2n}= pf (0,...,2n-1).$$
 whereas the $q_k$, defined in (3.3) from the
$p$'s, form an orthonormal sequence.
The matrix $Q$
defined by $q(z)=Q\chi(z)$ is the unique solution
(modulo signs)
 to the skew-Borel decomposition of $m_{\iy}$:
\be
m_{\iy}(t)=Q^{-1}JQ^{\top -1}, ~~\mbox{with}~~Q\in \Bk. \ee The
matrix $L=Q\Lb Q^{-1}$, also defined by $$z q(t,z)=L q(t,z),$$
 and the diagonal matrix $\tilde h$ satisfy the equations
\be
\frac{\pl L}{\pl t_i}=\left[-\pi_{\Bk}L^i, L \right].
~~\mbox{and}~~
\tilde h^{-1}\frac{\pl\tilde h}{\pl t_i}
=2\pi_{\Bk} (L^i)_0  .
\ee
\end{theorem}

\remark It is of interest to mention that, except for the
second expression for $p_{2n+1}$ in (3.7), the first part of
theorem 3.1 is actually {\em $t$-independent}. Namely, (3.7) gives
a geometric map, but also an effective algorithm, to perform the
skew-Borel decomposition of a skew-symmetric matrix $m_{\iy}$, by
applying formulae (3.7).

\bigbreak

\begin{corollary}
The polynomials $(p_0(t,z),p_1(t,z),...)$ have the following
explicit form in terms of Pfaffians of the moment matrix $m$ and
the Pfaffian $\TT$-functions:
\begin{eqnarray*}
 p_{2n}(z)&=&
\sum_{0\leq k\leq 2n}(-z)^k \frac{pf(0,...,\hat k,...,2n)}{pf(0,...,2n-1)}
\nonumber\\
&&\\
  p_{2n+1}(z)&=&
  z^{2n+1} +
 \sum_{0\leq k\leq 2n-1}(-z)^k  \frac{pf(0,...,\hat
k,...,\widehat{2n},2n+1) }{pf(0,...,2n-1)}.\\
\end{eqnarray*}
\be
\ee
\end{corollary}

\remark Taking into account the identity $pf($odd set$)=0$ and
the convention $$pf(0,...,\widehat
k,...,\widehat{2n},2n+1)=-pf(0,...,2n-1) ~~\mbox{ for}~~ k=2n+1, $$
the polynomials can be written as follows:
\bea
 p_{2n}(t,z)&=& \frac{1}{\TT_{2n}(t)}\sum_{0\leq k\leq
2n}(-z)^k pf(0,...,\hat k,...,2n,\widehat{2n+1})\nonumber\\
&&\nonumber\\
  p_{2n+1}(t,z)&=&
 \frac{1}{\TT_{2n}(t)}\sum_{0\leq k\leq 2n+1}(-z)^k pf(0,...,\hat
k,...,\widehat{2n},2n+1). \nonumber\\
\eea

 The proof of this theorem hinges on a number of lemmas,
involving properties of Pfaffians:

\begin{lemma} Consider an arbitrary skew-symmetric
 matrix $a=(a_{ij})_{i,j\geq 0}$. For
 odd $\ell$ and fixed $i\geq \ell$, the following holds:
$$
\sum_{0\leq k\leq\ell-1}(-1)^k a_{ ki}
pf(0,...,\hat k,...,\ell-1)=pf(0,...,\ell-1,i).
$$
\end{lemma}

\proof On the one hand
\be
\det(a_{ij})_{0\leq i,j\leq\ell}=pf^2(0,1,...,\ell)
\ee
and, on the other hand, using
 $$
\det \pmatrix{0& ... & a_{0r}&\Bigr| &a_{0k}\cr
          \vdots  &      &\vdots& \Bigr|& \vdots\cr
          -a_{0r} & ... & 0&\Bigr| &a_{rk}\cr
          --&--&--&--&--\cr
           a_{\ell 0} & ... & a_{\ell r}&\Bigr| &*
           }=pf(0,...,r,k) pf(0,...,r,{\ell}),
 $$
 and expanding according to the last row
and column

\bigbreak
\noindent$\displaystyle{\det(a_{ij})_{0\leq
i,j\leq\ell}}$
\bea
&=&-\sum_{0\leq i,j\leq\ell-1}(-1)^{i+j}a_{\ell
i}a_{j\ell}pf(0,...,\hat i,...,\ell-1)pf(0,...,\hat
j,...,\ell-1)\nonumber\\
 &=&\left(\sum_{0\leq i\leq\ell-1}(-1)^i
a_{\ell i}pf(0,...,\hat i,...,\ell-1)\right)\nonumber\\ &&
\hspace{5cm} \left(\sum_{0\leq j\leq\ell-1}(-1)^j a_{\ell j}pf(0,...,\hat
j,...,\ell-1)\right)\nonumber\\
 &=&\left(\sum_{0\leq
i\leq\ell-1}(-1)^i a_{i\ell }pf(0,...,\hat
i,...,\ell-1)\right)^2\quad.
\eea
Comparing the two expressions (3.12) and (3.13) and taking into
account the sign, lead to the claim of lemma 3.3.\qed

\remark Lemma 3.3 will be applied as follows (for odd $\ell$):
$$
\sum_{0\leq k\leq\ell-1}(-1)^k a_{ i_k,i_{\ell}}
pf(i_0,...,\hat
i_k,...,i_{\ell-1})=pf(i_0,...,i_{\ell-1},i_{\ell}).
$$

\vspace{0.2cm}

\begin{lemma} The polynomials $p_k(z)$ defined by the determinantal expression (3.7) equal the
 expressions (3.10) of Corollary 3.2 and
satisfy
\be \left\la p_i(z),p_j(z)\right\ra_{i,j\geq 0}=J \tilde h
\ee
\end{lemma}

\proof At first notice that the leading coefficients of the
pfaffians in the polynomials (3.7) equal $pf(0,1,...,2n-1)=\tilde
\tau_{2n}(t)$. Then apply Lemma 3.3 (remark following the lemma)
 to the first case, setting
the $a$'s equal to $\mu$'s from $0$ to $2n$ and
$a_{k,2n+1}=(1-\dt_{k,2n+1})z^k$. Also apply Lemma 3.3 to the
second case, by setting the $a$'s equal to $\mu$, from $0$ to
$2n+1$, skipping $2n$, and $a_{k,2n}=(1-\dt_{k,2n})z^k$. Then, we
find:
\bea
\TT_{2n}(t)  p_{2n}(t,z)&=& \sum_{0\leq k\leq
2n}(-z)^k pf(0,...,\hat k,...,2n-1,2n,\widehat{2n+1})\nonumber\\
 &=&
\sum_{0\leq k\leq 2n}(-z)^k pf(0,...,\hat
k,...,2n)\nonumber\\
 &&\nonumber\\
 \TT_{2n}(t) p_{2n+1}(t,z)&=&
 \sum_{0\leq k\leq 2n+1}(-z)^k pf(0,...,\hat
k,...,2n-1,\widehat{2n},2n+1) \nonumber\\
 &=&
 \sum_{0\leq k\leq 2n-1}(-z)^k pf(0,...,\hat
k,...,\widehat{2n},2n+1) \nonumber\\
 && ~~~~~+ z^{2n+1} pf(0,...,2n-1) .\nonumber\\
\eea
Using the expressions (3.15) and Lemma 3.3, one computes the
inner-product
\begin{eqnarray*}
\la p_{2n}(z),z^i\ra&=&\sum_{0\leq k\leq
2n}(-1)^k\mu_{ki}\frac{pf(0,...,\hat k,...,2n)}{pf(0,...,2n-1)}\\
&=&\frac{pf(0,...,2n,i)}{pf(0,...,2n-1)},
\end{eqnarray*}
which vanishes for $0\leq i\leq 2n$, and so
$$
\la p_{2n}(z),z^i\ra=0 ~~\mbox{for}~~0\leq i \leq 2n;
$$
therefore
$$
\la p_{2n}(z),p_i(z)\ra=0
 ~~\mbox{for}~~0\leq i \leq 2n.
$$
Moreover, using the expression for $p_{2n+1}$ and Lemma 3.3,
\begin{eqnarray*}
\la p_{2n+1}(z),z^i\ra&=&\mu_{2n+1,i} +\sum_{0\leq k\leq
2n-1}(-1)^k\mu_{ki}\frac{pf(0,...,\hat
k,...,\widehat{2n},...,2n+1)}{pf(0,...,2n-1)}\\ & &\quad\quad
\\ &=&\frac{pf(0,...,\widehat{2n},2n+1,i)}{pf(0,...,2n-1)}\quad
\\ &=&\left\{\begin{array}{l}
0\mbox{\,\,for\,\,}0\leq i\leq 2n+1,i\neq 2n\\
 \\
\frac{pf(0,...,\widehat{2n},2n+1,2n)}{pf(0,...,2n-1)}
=-\frac{p(0,...,2n,2n+1)}{pf(0,...,2n-1)}
=-\frac{\TT_{2n+2}(t)}{\TT_{2n}(t)}=-\tilde h_{2n}\\
\hspace{7.5cm}\mbox{\,\,for\,\,} i=2n.
\end{array}\right.
\end{eqnarray*}
and so
$$
\la p_{2n+1}(z),p_i(z)\ra=-\dt_{i,2n}\tilde h_{2n}
 ~~\mbox{for}~~0\leq i \leq 2n+1;
$$
the skew-symmetry of $\la,\ra$ does the rest. \qed

\begin{lemma} We have
$$
\frac{\pl}{\pl t_k}pf(i_0,...,i_{\ell})=\sum^{\ell}_{r=0}
pf(i_0,...,i_r+k,...,i_{\ell}).
$$
\end{lemma}

\proof The differential equations (2.1) for $m_{\iy}(t)$ read
in terms of the elements $\mu_{ij}$:
\be
\frac{\pl \mu_{ij}}{\pl t_k}=\mu_{i+k,j}+\mu_{i,j+k};
\ee
hence, setting $\sigma (i)=i'$, we compute

\noindent $\displaystyle{\frac{\pl}{\pl t_k}pf(i_0,...,i_{n-1})}$
\begin{eqnarray*}
&=&\frac{1}{(n/2)!
2^{n/2}}\sum_{\sg}\vr(\sg)\sum_{{r\in\{0,...,n-2\} }\atop {r
~\mbox{\tiny{even}}}}\mu_{i'_0i'_1}...
(\mu_{i'_r+k,i'_{r+1}}+\mu_{i'_r,i'_{r+1}+k})...\mu_{i'_{n-2},i'_{n-1}}\\
\\
&=&\sum_{r=0}^{n-1}pf(i_0,...,i_{r}+k,...,i_{n-1}),
\end{eqnarray*}
upon reorganizing the first sum on the right hand side
appropriately.\qed

\begin{lemma}
The polynomials $p_k$ of Theorem 3.1 satisfy:
$$
\frac{1}{\TT_{2n}(t)}\left(\frac{\pl}{\pl t_1}+z\right)
\TT_{2n}(t)p_{2n}(z)=p_{2n+1}(z).
$$
\end{lemma}

\proof
Using the explicit form (3.10) of $p_{2n}$ and Lemma 3.5, one
computes

\medbreak
\noindent$\displaystyle{
\left(z+\frac{\pl}{\pl t_1}
\right)\TT_{2n}(t)p_{2n}(z)}$
\begin{eqnarray*}
&=&-\sum_{k=0}^{2n}(-z)^{k+1}pf(0,1,...,\hat k,...,2n)\\ &
&\quad\quad
+\sum_{k=0}^{2n}(-z)^k(pf(0,...,\widehat{k-1},...,2n)+pf(0,...,\hat
k,...,\widehat{2n},2n+1))\\ &=&z^{2n+1}pf(0,1,...,2n-1)+z^{2n}
\left(-pf(0,...,\widehat{2n-1},2n)+
pf(0,...,\widehat{2n-1}, 2n)\right)\\ & &\quad +\sum_1^{2n-1}(-z)^k
pf(0,...,\hat k,...,\widehat{2n},2n+1)+pf(1,...,\widehat{2n},
2n+1)\\ &=&
\TT_{2n}(t)p_{2n+1}(z),
\end{eqnarray*}
ending the proof.\qed

\begin{lemma}
The decomposition
\be
m_{\iy}=Q^{-1}J Q^{\top -1},~~\mbox{with}~~Q\in \GR_{\Bk},
\ee
is unique (modulo signs) and is completely given by the
skew-orthonormal polynomials $q_k(z)$.
\end{lemma}

\proof In view of relations (3.5), the skew-orthonormal polynomials
(3.3) provide a solution to the Borel decomposition
$m_{\iy}=Q^{-1}J Q^{\top -1}$, with $Q\in \GR_{\Bk}$. We now show
this $Q\in
\GR_{\Bk}$ is unique! To do so, we proceed by induction on
$q_k=(Q\chi(z))_k$. Assuming the existence of another matrix
$\tilde Q \in \GR_{\Bk}$, such that $m_{\iy}=\tilde Q^{-1}J
\tilde Q^{\top
-1}$, we must show $q=\tilde q$. Indeed $m_{\iy}=Q^{-1}J Q^{\top -1}$
leads unambiguously (except for the sign) to
$$
q_0=\mu_{01}^{-1/2},~~\mbox{and}~~q_1=\mu_{01}^{-1/2}z.
$$
But, for some $\al_i$ and $\beta_i$, we would have, taking into
account the absence of the $z^{2n}$-term in $\tilde q_{2n+1}$,
\be
\tilde q_{2n}=\sum_0^{2n} \al_i q_i~~\mbox{and}~~
\tilde q_{2n+1}=\al_{2n}q_{2n+1}+\sum_0^{2n-1} \beta_i q_i.
\ee
Then using the inductive step, namely $\tilde q_{k}=q_{k}$ for $0
\leq k \leq 2n-1$, and the skew-orthonormality, we compute for $\ell < n$,
\bea
0&=&\la \tilde q_{2n} ,\tilde q_{2\ell} \ra=\la \tilde q_{2n} ,
q_{2\ell} \ra=-\al_{2\ell+1}\nonumber\\
 0&=&\la \tilde q_{2n},\tilde q_{2\ell +1} \ra=
 \la \tilde q_{2n}, q_{2\ell +1} \ra =\al_{2\ell}\nonumber\\
0&=&\la \tilde q_{2n+1},\tilde q_{2\ell} \ra =\la \tilde q_{2n+1},
q_{2\ell} \ra =-\beta_{2\ell+1}\nonumber\\ 0&=&\la
\tilde q_{2n+1},\tilde q_{2\ell+1}\ra=\la
\tilde q_{2n+1}, q_{2\ell+1} \ra=\beta_{2\ell},
\nonumber
\eea
and so, since also $\beta_{2n}=0$,
$$
\tilde q_{2n}=\al_{2n}  q_{2n}~~\mbox{and}~~\tilde
 q_{2n+1}=\al_{2n}  q_{2n+1}.
 $$
 Finally
 $$
 1=\la \tilde q_{2n}, \tilde q_{2n+1} \ra
 =\al_{2n}^2 \la  q_{2n}, q_{2n+1} \ra = \al_{2n}^2,
 $$
leading to  $\al_{2n}^2=1$. \qed

\underline{\sl Proof of Theorem 3.1 and Corollay 3.2}: The proof
follows at once from lemmas 3.3 up to 3.7. The fact that $Q$ and
$L$ satisfy the differential equations (2.3) and (2.4) follows from
the fact that the matrix $m_{\iy}$ satisfies the differential
equations
$$\pl m / \pl t_k=\Lb^k m
+m \Lb^{\top k}, $$ according to Theorem 2.1.

The second relation in (3.9) is established as follows: defining
the semi-infinite lower-triangular matrix $P$ (with $1$'s along the
diagonal) by means of the monic skew-orthogonal polynomials $p(z)=P
\chi(z)$ and differentiating $P=\tilde h^{1/2} Q$ with regard to
$t_k$, one
 obtains
$$
\tilde h^{1/2} \dot Q Q^{-1}\tilde h^{-1/2}=
-\frac{1}{2}(\log \tilde h)^. +  \dot P P^{-1}.
$$
Then, using proposition 2.2, (formula (II)), one finds:
\begin{eqnarray*}
0=\tilde h^{1/2}(\mbox{II})_0 \tilde h^{-1/2}&=&
 \tilde h^{1/2}\left((L^k)_0-J(L^{k
\top})_0 J+2(\dot Q Q^{-1})_0\right)\tilde h^{-1/2}\\
&=&\tilde h^{1/2}2 \pi_{\Bk}(L^k)_0\tilde h^{-1/2}-
  (\log \tilde h)^. +2(\dot P~P^{-1})_0\\
&=& 2 \pi_{\Bk}(L^k)_0 -
  (\log \tilde h)^.~,
\end{eqnarray*}
using the commutation of $\pi_{\Bk}(L^k)_0$ with $\tilde h^{1/2}\in
\DR_0^-$, together with $(\dot P~P^{-1})_0=0$.
\qed

\end{document}